\documentclass[aps,prl,twocolumn,showpacs,floatfix,a4paper]{revtex4}
\usepackage{graphicx}
\usepackage[intlimits]{amsmath}

\begin{document}

\title{Entanglement entropy in one-dimensional disordered interacting 
system: The role of localization}
\author{Richard Berkovits}
\affiliation{Department of Physics, Bar-Ilan
University, Ramat-Gan 52900, Israel}

\begin{abstract}
The properties of the entanglement entropy (EE) in one-dimensional
disordered interacting systems are studied. Anderson localization
leaves a clear signature on the average EE, as it saturates 
on length scale exceeding the localization length. This is
verified by numerically calculating the EE for an ensemble of disordered
realizations using density matrix renormalization group (DMRG).  
A heuristic expression describing the dependence of the EE on the
localization length, which takes into account finite size effects, is
proposed. This is used to extract the localization
length as function of the interaction strength. 
The localization length dependence on
the interaction fits nicely with the expectations.
\end{abstract}

\pacs{73.20.Fz,03.65.Ud,71.10.Pm,73.21.Hb}

\maketitle

\emph{Introduction.}---
Although more than half a century has passed since
the seminal work of Anderson on the absence of diffusion
in certain random lattices \cite{anderson58},
the localization transition, with an emphasis on 
understanding its behavior in strongly correlated systems,
remains one of the central themes of condensed
matter physics. Unlike many other quantum phase transitions
in which a gap between the ground state and the excited state
energies appear, the localization transition is more subtle. The
canonical manifestation of the localized phase is the exponential
dependence of the conductance, $G$, on the linear
dimension of the system, $L$. The localization length, $\xi$, 
(corresponding
to the correlation length in the localized regime) is defined
through the exponential decrease in the conductance 
$G(L) \sim \exp(-L/\xi)$ \cite{lee85}.

For non-interacting systems the localization length can
be visualized easily as the the length-scale on which
the single-electrons wave-function envelope decays.
No such transparent interpretation exists for the many-particle
wave-function. Indeed, the numerical extraction of the localization 
length for interacting systems is difficult. Direct calculation
of the conductance is computationally taxing since it demands
knowledge of the excited states. Ground state properties
are easier to calculate, but there one runs into
other problems. For example, using  the sensitivity to
boundary conditions \cite{schmitt98} 
measured, e.g., by the persistent current, 
depends both on the localization
length as well as on the inverse
compressibility of the system \cite{berkovits96}. Calculating the two 
requires both a calculation of
the sensitivity of the ground state to a magnetic flux in periodic 
boundary conditions as well as the dependence
of the number of electrons in the system on the
chemical potential. 
Alternatively, one may study the response of the system
to the introduction of a perturbation \cite{carter05}. For example,
the decay of the
Friedel oscillations resulting from the introduction of an additional
impurity to the system \cite{weiss07}. Nevertheless, 
this is strictly speaking justified only
for weak disorder, and one must calculate the ground state both in the presence
and absence of the impurity. 

In this letter we will utilize the entanglement entropy \cite{amico08}
(EE, sometimes referred to as the von Neumann or Shanon entropy) in order to 
extract the localization length of a 1D interacting disordered system.
EE for a many-body system in a pure state is related to
partitioning it into two regions: A and B. The
entanglement between A and B is measured by the EE $S_{A/B}$
in the following way: Using the Schmidt decomposition the
many-body pure state of the entire system, $|\Psi\rangle$,
is expressed as the sum of two orthonormal
basis sets of regions A ($\{|\phi_{A,i}\rangle\}$) and B
($\{|\phi_{B,j}\rangle\}$), in the following way
\begin{eqnarray}
|\Psi\rangle=\Sigma_{i}\alpha_{i} |\phi_{A,i}\rangle \otimes |\phi_{B,i}\rangle,
\label{phi}
\end{eqnarray}
with $\Sigma_{i}\alpha_{i}^2=1$.

The Schmidt decomposition is related to the eigenbasis of the 
reduced density operators
$\hat \rho_{A/B}={\rm Tr}_{B/A}|\Psi\rangle\langle \Psi |$, by
\begin{eqnarray}
\hat \rho_{A/B}=\Sigma_{i}\alpha_{i}^2|\phi_{A/B,i}\rangle \langle \phi_{A/B,i}|.
\label{e3}
\end{eqnarray}
The EE measures
the entanglement between the two regions by the
von Neuman entropy of the reduced density matrix:
\begin{eqnarray}
S_{A/B}=-\Sigma_{i}\alpha_{i}^2\ln(\alpha_i^2), 
\label{e4}
\end{eqnarray}
equivalent to the Shannon entropy of the squared 
Schmidt coefficients $\alpha_i^2$, clearly $S_A=S_B$. 

The behavior of the EE 
has recently been connected 
to the presence of quantum phase transitions (QPTs)
in condensed mater systems \cite{amico08,vojta06,lehur08,goldstein11}.
Specifically, the EE of a region of length $L_A$ of a
1D metallic system 
is expected to grow logarithmically with $L_A$
\cite{holzhey94,vidal03,calabrese04},
while the EE in the insulating regime should not depend on the regions size.
Since for length scales shorter than the localization length the system
behaves essentially as a metal, we expect that the EE of a disordered
1D system will show a logarithmic dependence for $L_A \ll \xi$, while it
will saturate for $L_A \gg \xi$. Tracing the crossover will be used to
determine the localization length $\xi$, even for an many-particle interacting
ground state.

Density matrix renormalization group (DMRG) \cite{white92,dmrg}
is the natural choice for a numerical method to calculate the ground state 
properties of disordered interacting 1D system. Moreover, since the Schmidt
coefficients are calculated in the DMRG in order to decide which states
should be truncated, calculating $S$ has no additional computational
overhead. Since DMRG is much more accurate for finite systems with
open boundaries than 
for periodic boundary conditions (another drawback for methods
relaying on persistent current calculations), we prefer to consider
a finite system of size $L$ in which the EE of a region of length $L_A$
starting at the edge is calculated. Thus, the EE will depend both on
$L_A$ as well as on $L$. The explicit function $S(L_A,L)$ as been
derived by several authors \cite{holzhey94,vidal03,calabrese04,korepin04},
and for open boundaries:
\begin{equation}
S(L_A,L)=\frac{1}{6}\ln\left(\sin\left(\frac{\pi L_A}{L}\right)\right)+c,
\label{sin}
\end{equation}
where $c$ is a non-universal constant which depends also on $L$ and the 
boundary entropy \cite{affleck91}. This dependence was tested numerically
for 1D spin chains \cite{igoli08}, and found to fit pretty well.

\emph{Model.}---
We consider a spinless 1D electrons wire of size $L$ 
with repulsive nearest neighbor (NN)
interactions and on-site disorder. 
The system's Hamiltonian is thus given by the Anderson model
\begin{eqnarray} \label{hamiltonian}
H &=& 
\displaystyle \sum_{j=1}^{L} \epsilon_j {\hat c}^{\dagger}_{j}{\hat c}_{j}
-t \displaystyle \sum_{j=1}^{L-1}({\hat c}^{\dagger}_{j}{\hat c}_{j+1} + h.c.) \\ \nonumber
&+& U \displaystyle \sum_{j=1}^{L-1}({\hat c}^{\dagger}_{j}{\hat c}_{j} - \frac{1}{2})
({\hat c}^{\dagger}_{j+1}{\hat c}_{j+1} - \frac{1}{2}),
\end{eqnarray}
where $\epsilon_j$ is the random on-site energy, taken from a uniform 
distribution in the range $[-W/2,W/2]$,
$U$ is the NN interaction strength ($U \ge 0$), and $t=1$ is the
hopping matrix element between NN.
${\hat c}_j^{\dagger}$ is the creation 
operator of a spinless electron at site $j$ in the wire, and
a positive background is included in the interaction term.
For the non-interacting case the dependence of the localization
length on the disorder is known $\xi(W,U=0) \approx 105/W^2$
\cite{romer97}. 

For the interacting case, 
using renormalization group \cite{apel_82} the following
dependence of the localization length on interaction strength and
disorder is suggested
\begin{eqnarray} \label{xi_u}
\xi(W,U) = (\xi(W,U=0))^{1/(3-2g(U))},
\end{eqnarray} 
where $g(U)=\pi/[2 \cos ^{-1} (-U/2)]$ 
is the Luttinger parameter \cite{g_formula}. 
For non-interacting electrons $g(U=0)=1$.
Since for repulsive interactions $g<1$ decreases as a 
function of the interaction strength, one finds that the localization length 
always decreases as a function of the interaction strength 
\cite{schmitt98,giamarchi}.

\begin{figure}
\includegraphics[width=8.5cm,height=!]{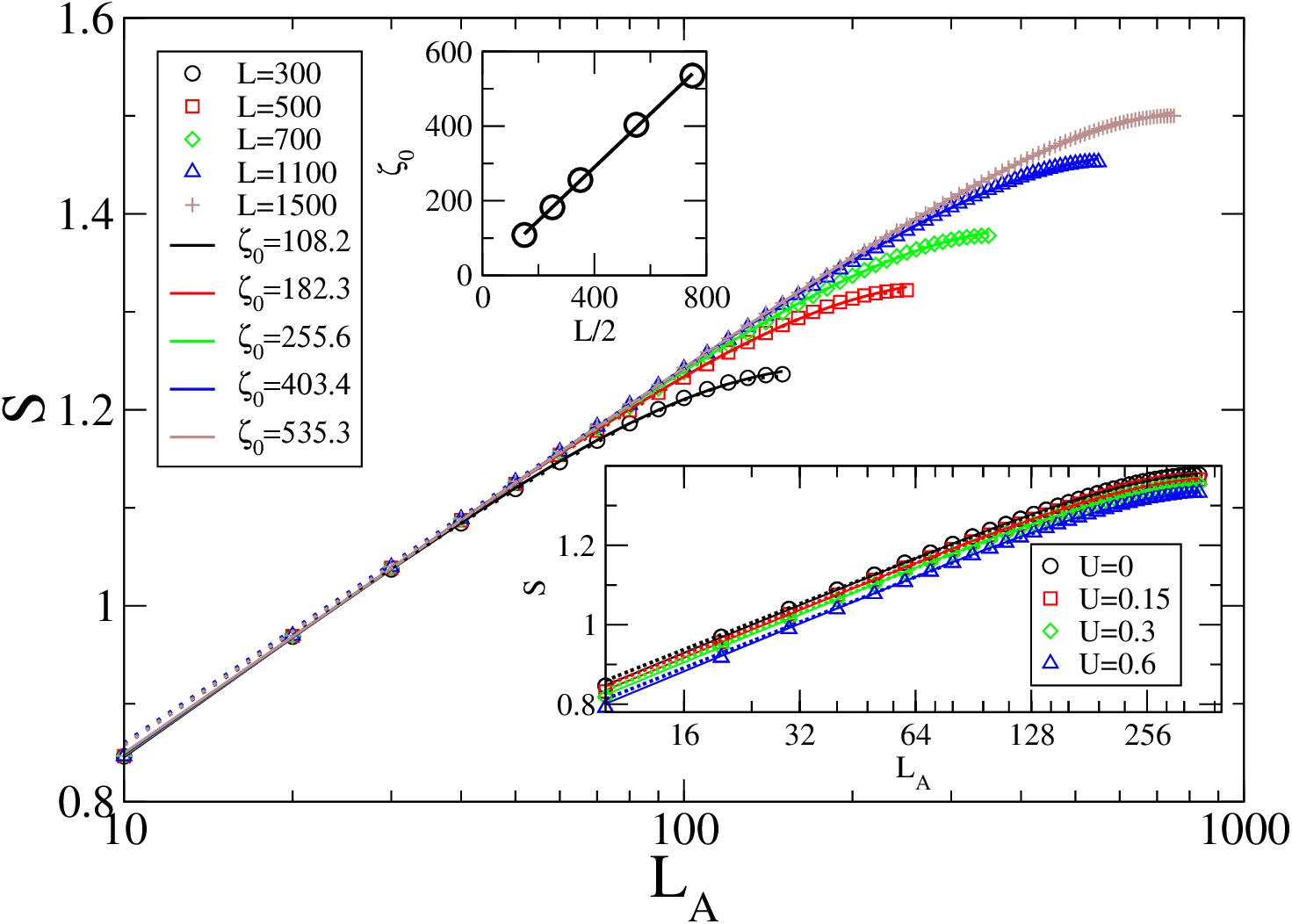}
\caption{\label{fig:clean}
(Color online)
The EE of a clean system as function of $L_A$,
for different system sizes $L$.
The symbols correspond to the DMRG results, the dotted curves to Eq. (\ref{sin})
and the continuous curves to Eq. (\ref{ee_erf}), 
with a value of $\zeta_0$ indicated in the legend. 
Bottom inset:
The EE of a clean system 
for different values of interaction strength $U$ and $L=700$.
The symbols correspond to the DMRG results, the dotted curves to Eq. (\ref{sin})
and the continuous curves to Eq. (\ref{ee_erf}), 
with $\zeta_0=255.6$. 
Top inset: $\zeta_0$ as function of $L$. The curve corresponds to 
$\zeta_0(L)=0.357 L$. 
}
\end{figure}

\emph{Clean systems.}---
We begin by calculating the EE for clean systems of different
sizes as function of the length of region A
($L_A$). During the DMRG calculations $192$ states are retained. The accuracy
may be assessed by comparing $S(L_A,L)$ to $S(L-L_A,L)$.
The results are presented in Fig. \ref{fig:clean}. The EE 
for all system length begin with a logarithmic dependence on $L_A$ for
$L_A \ll L$ and tapers out for $L_A \approx L/2$. Fitting the numerical
results to Eq. (\ref{sin}), where the only fitting parameter is the 
constant $c$, works quite well, except for some deviations at small
values of $L_A$. Such deviations were also seen for spin chains with open
boundary conditions (see Ref. \onlinecite{igoli08}).

Adding interactions ($U \ne 0$) to a clean system
does not change the conductance and the system remains metallic. Thus,
one expects that the EE will behave in a similar way in the presence or
absence of interactions. Indeed, this can be seen in the bottom
inset of Fig. \ref{fig:clean},
where the EE as function of $L_A$ is calculated for different interaction
strength ($U=0,0.15,0.3,0.6$ corresponding to $g=1,0.954,0.913,0.836$),
for $L=700$. Except for a small change in the constant $c$, the
EE shows the same functional behavior for the different values of interaction.


\emph{Disorder.}--- 
The study of the influence of disorder on the EE of 1D 
systems has concentrated on the influence of a 
single impurity in the middle of the sample
\cite{zhao06,levine04,peschel05,igloi09,ren09,affleck09,eisler10}.
For the non-interacting Anderson model \cite{eisler10} 
the system remains metallic even in the presence of an impurity and the
EE continues to be described by a logarithm, though with a different slope. 
For the interacting case, (i.e. for a Luttinger liquid) 
the impurity effectively cuts the system into two, driving the
system into an insulating phase \cite{kane92}, resulting in a saturation
of the EE \cite{zhao06}. The multiple impurity situation was only
discussed for the non-interacting case \cite{jia08,chakravarty10}, 
where the EE of a single cite (i.e., $L_A=1$) was shown to correspond
to the well known measure of the inverse participation ratio. 

When disorder is introduced, the main changes
anticipated in the behavior of $S(L_A,L)$ are the saturation of
of the EE for $L>L_A\gg\xi$ and a metallic behavior (i.e., logarithmic,
however with a non-universal slope) for $L,\xi \gg L_A$. 
The following considerations are taken into account while 
suggesting a functional
description of the EE: 
For small values of $x=L_A/{\rm min}\{\xi,L/2\}$
one expects
$S(L_A,\xi,L) \approx S(x) \approx \ln(x)$ 
while for large values
of $x$, $ S(x) \approx {\rm Constant}$. Thus, it is natural
to assume that $S(x)$ takes the form $S(x)\approx \ln(f(x))$, where
$f(x) \approx x$ for $x \ll 1$ and $f(x) \approx {\rm Constant}$ for $x\gg 1$.
The most natural choice
would be $f(x)=\tanh(x)$, nevertheless it turns out that
the crossover between the two regions ($f(x) \approx x$ and $f(x) 
\approx {\rm Constant}$) is quite sharp and $\tanh(x)$ does not describe it
well. The straight forward way to
sharpen the transition would be to replace the exponential governing
$\tanh(x)$ by a stronger decaying function, e.g. a Gaussian,
hence $f(x) = {\rm erf}(x)$
(the error function) which provides a better description.
Thus, we propose to fit the EE to the
following heuristic expression:

\begin{eqnarray} \label{ee_erf}
S(L_A,\xi,L) = s(\xi) \ln\left({\rm erf}\left(\frac{L_A}{\zeta(L,\xi)}\right)
\right)+c,
\end{eqnarray} 
where $s(\xi)<1/6$, is a constant which depends on the disorder,
$\zeta(L,\xi)$ depends both on $L$ and $\xi$ and $c$ is a non-universal 
constant. It is important to note
that for the disordered case one discusses a disorder averaged 
value of the EE.

For the clean limit
($\xi \rightarrow \infty$), we may infer the behavior of
$\zeta_0(L)=\zeta(L,\infty)$
from fitting Eq. (\ref{ee_erf}) to the data we obtained for different
system length presented in Fig. \ref{fig:clean}.
The fit is represented by the continuous curves in Fig. \ref{fig:clean}.
As expected $s(\xi \rightarrow \infty) \approx 1/6$, while the results
for $\zeta_0(L)$ are depicted in the legend.
One can see that the heuristic expression for the EE (Eq. (\ref{ee_erf}))
provides a very good description of the numerical data with an even
somewhat better fit than Eq. (\ref{sin}) for small values of $L_A$.
In the top inset of Fig. \ref{fig:clean} $\zeta_0$ as function of
$L$ is presented. A linear relation $\zeta_0=0.357L$ emerges.
Eq. (\ref{ee_erf}) fits also the behavior of the EE for the 
clean interacting case (bottom inset Fig. \ref{fig:clean}) where only $c$
depends on the interaction strength. 

\begin{figure}
\includegraphics[width=8.5cm,height=!]{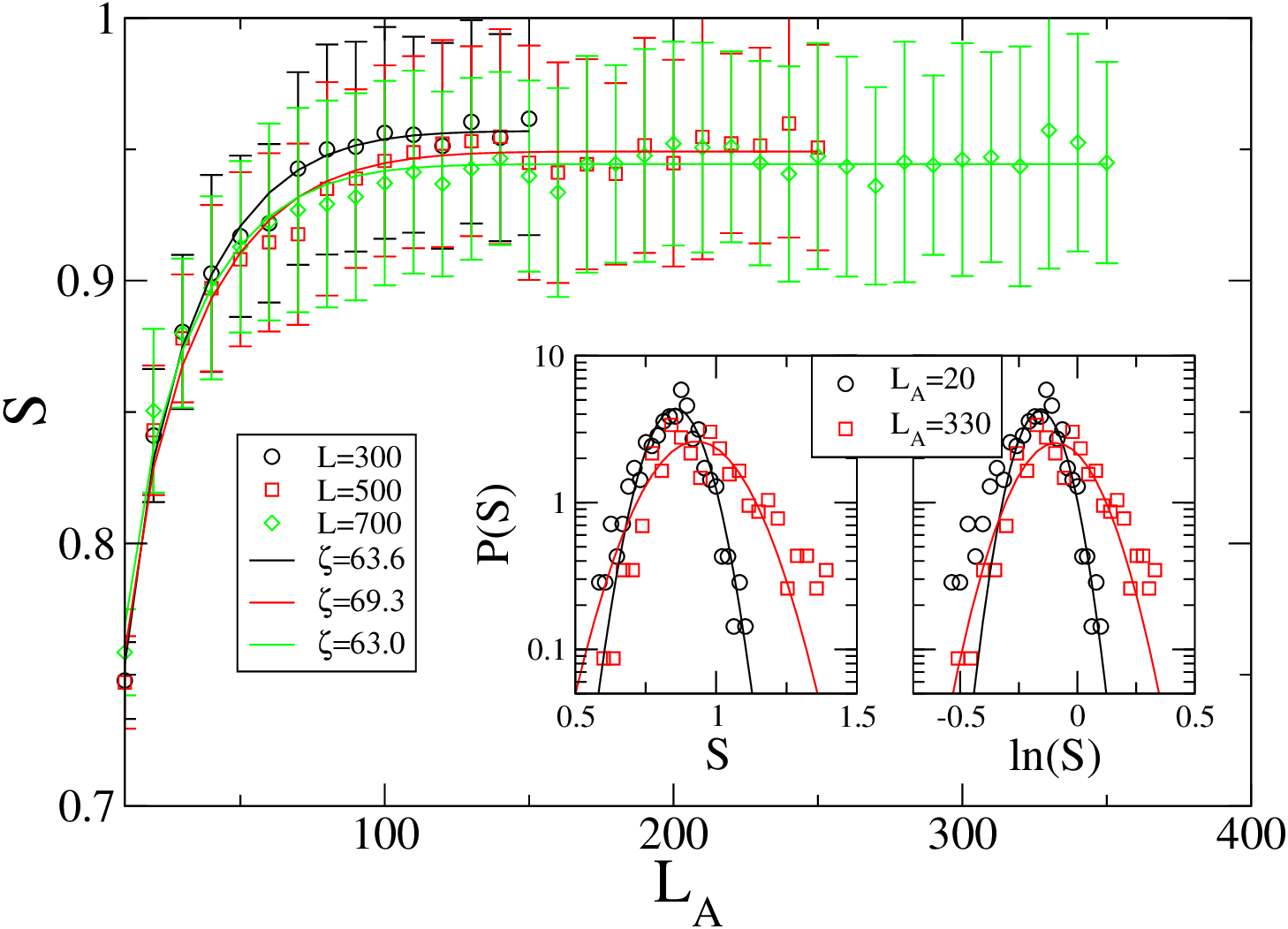}
\caption{\label{fig:u06}
(Color online)
The median of the EE ($S(L_A,\xi,L)$) as function of $L_A$ 
for system length $L=300,500,700$. The median was calculated for
ensembles of $600,400,400$ correspondingly. In all cases,
the disorder $W=0.7$ ($\xi=214$) and the interaction
strength $U=0.6$ ($g=0.836$). 
Error bar represent range between the 40th and 60th percentiles.
The curves represent a fit to Eq. (\ref{ee_erf}) with values of
$\zeta$ indicated in the legend.
Left inset: The probability distribution of the EE for $L=700$
and $L_A=20,330$. The symbols represent the numerical data,
while the curves correspond to a fit to the normal distribution.
Right inset: The same numerical data as in the left inset, fitted to
a log-normal distribution. 
}
\end{figure}

In the limit of $\xi\ll L$ one would expect that the system size $L$
will have only a rudimentary influence on $\zeta(L,\xi)$, i.e.,
$\zeta(L\gg\xi) \approx \xi$. We investigate this assertion by
studying the behavior of $S(L_A,\xi,L)$ for $W=0.7$ and $U=0.6$
for $L>300$. Using 
Eq. (\ref{xi_u}), one expects that the localization length $\xi = 57$, 
is much smaller than $L$. 
Prior to examining the behavior of the average value $S(L_A,\xi,L)$,
it is important to examine the behavior of the distribution of the EE
for different realizations, in order to verify that the concept of an
average value of the EE is meaningful, especially since we are in the
localized regime. The distribution of $S(L_A,\xi,L)$
for $L=700$ derived from an ensemble of $400$ realizations 
and different values of $L_A$ is presented in the insets
of Fig. \ref{fig:u06}. In the left inset the distribution is fitted to
a normal distribution while in the right inset to a log-normal distribution.
While for $L_A=20$ the distribution follows a normal distribution
for $L_A=330$ a fit to the log-normal distribution works better.
Thus, it seems, that while for $L_A\ll \xi$ the EE distribution is normal, 
for $L_A\gg \xi$ it is log-normal,
somewhat similar to the behavior of the distribution of the conductance
\cite{anderson80}. Thus, a better representation of the typical EE
is the median, which is less sensitive to the tails of the distribution. 
The errors are estimated
by the $40$ (and $60$) percentile of the distribution.

The median EE for samples of length $L=300,500,700$ over ensembles
of $600,400,400$ realizations 
correspondingly are presented in Fig. \ref{fig:u06}.
It is clear that the EE does not change much for the different length of
the samples. Fitting Eq. (\ref{ee_erf}) results in $\zeta = 63.6.69.6,63$
for $L=300,500,700$. This is in line with our expectations for $\zeta \approx
\xi$ for $L \gg \xi$.

\begin{figure}
\includegraphics[width=8.5cm,height=!]{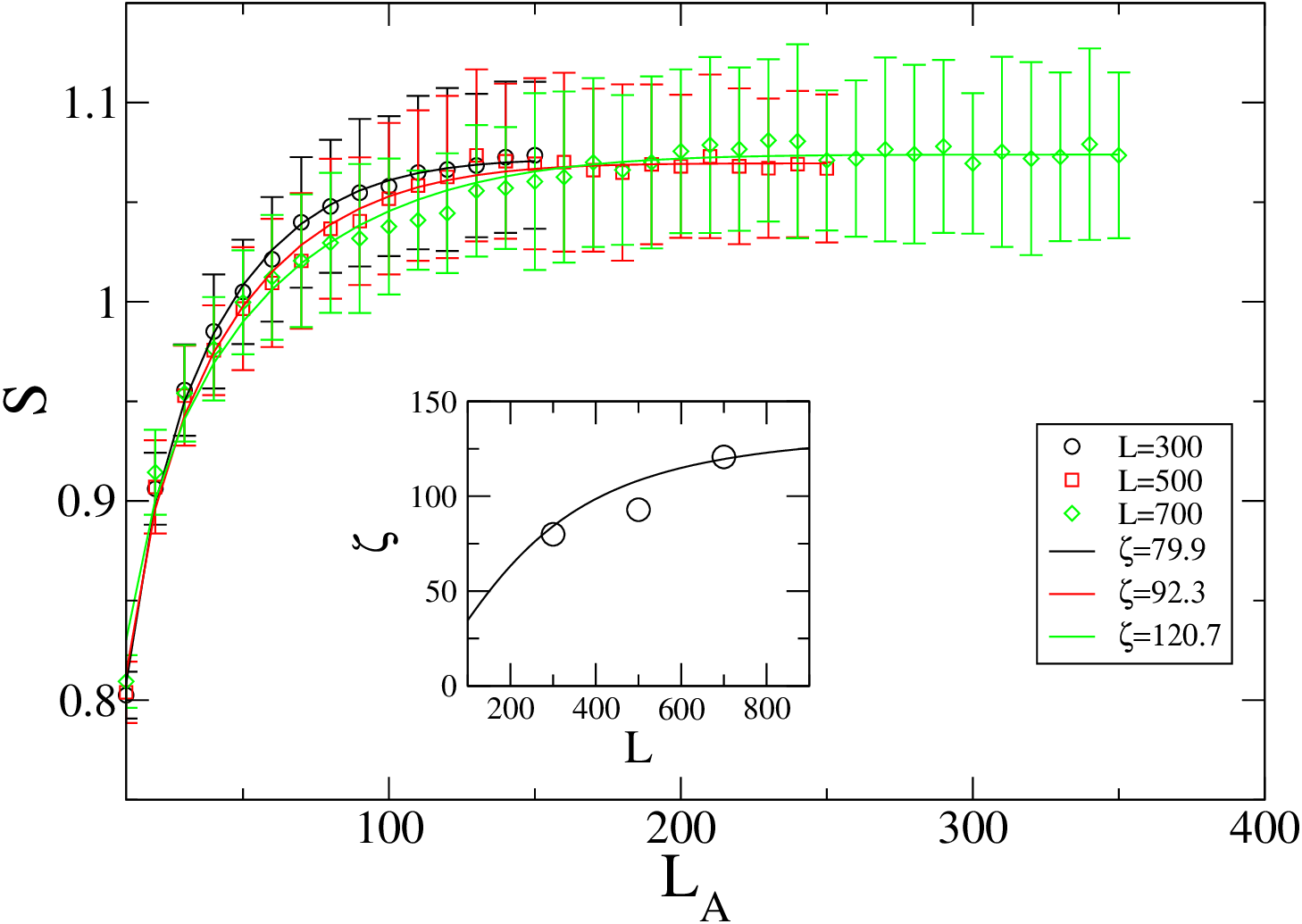}
\caption{\label{fig:u015}
(Color online)
As in Fig. \ref{fig:u06}, for weak interactions
($U=0.15$,$g=0.954$)). 
The curves represent a fit to Eq. (\ref{ee_erf}) with values of
$\zeta$ indicated in the legend.
Inset: Comparison of the interpolation formula proposed in Eq.
(\ref{zeta}) represented by the curve, and the numerical results for $\zeta$
(circles).
}
\end{figure}

For the case where both the finite-size of the sample and the localization
length are of comparable length, one expects that both will influence 
$\zeta(L,\xi)$. Since the limits are clear, i.e., 
$\zeta(L\gg \xi) \approx \xi$ and $\zeta(L\ll \xi) 
= \zeta_0=0.357 L$, one may interpolate between these two limits by:
\begin{eqnarray} \label{zeta}
\frac{1}{\zeta^2} = \frac{1}{\zeta^2_0}+\frac{1}{\xi^2}.
\end{eqnarray} 
In Fig. \ref{fig:u015} we examine the behavior of the EE for a weakly 
interacting
($U=0.15,g=0.954$) case. According to Eq. (\ref{xi_u}), $\xi=136$
is close to the values of $\zeta_0$ expected for the system sizes
$L=300,500,700$ examined. A fit of the numerical data
to Eq. (\ref{ee_erf}) results in $\zeta=79.9,92.3,120.7$ for $L=300,500,700$.
As can be seen in the inset, these values of $\zeta$ fit reasonably  
with the extrapolation of $\zeta$ obtained from Eq. (\ref{zeta}).

\begin{figure}
\includegraphics[width=8.5cm,height=!]{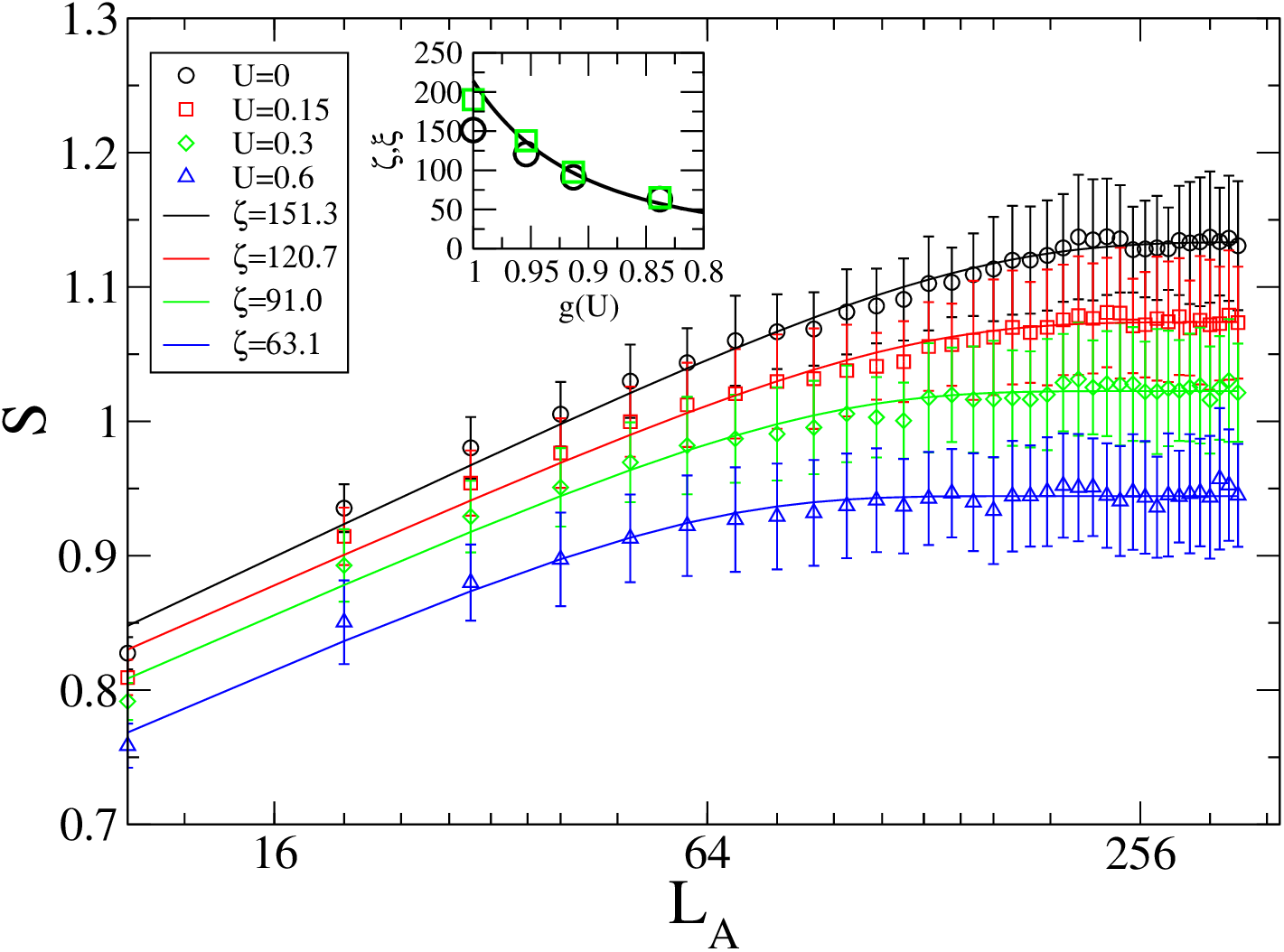}
\caption{\label{fig:u}
(Color online)
The median of the EE as function of $L_A$ for system length $L=700$ 
and $W=0.7$ ($\xi=214$) for different values of interaction
strength $U=0,0.15,0.3,0.6$ corresponding to $g=1,0.954,0.913,0.836$. 
The curves represent a fit to Eq. (\ref{ee_erf}) with values of
$\zeta$ indicated in the legend.
Inset: The values of $\zeta$ (depicted by circles) and
$\xi$ (extracted from Eq. (\ref{zeta}) using $\zeta_0=255.6$, 
indicated by squares) as function of the Luttinger parameter $g$.
The curve corresponds to the values of $\xi$ expected
according to Eq. (\ref{xi_u}).
}
\end{figure}

Now we examine the role played by interactions
on the behavior of the localization length. In Fig. \ref{fig:u}
we show the EE of the longest samples ($L=700$) for which 
it was feasible to obtain an adequate
number of realizations. As expected the saturation
of the EE begins for smaller values of $L_A$ as the interaction increases. 
A more quantitative comparison
of the behavior of the localization length may be achieved by fitting
the median EE to Eq. (\ref{ee_erf}) and extracting $\zeta$. Finite size
corrections (especially important when $\zeta$ is comparable
to $\zeta_0$, i.e. for weak interactions) can be taken into
account by the extrapolation suggested in Eq. (\ref{zeta}).
The localization length thus obtained is compared to the expectations
according  to Eq. (\ref{xi_u}) in the inset. We see that the localization
length obtained from the EE fits quite well the expectations for the
dependence of $\xi$ on the interaction strength.

\emph{Conclusions.}---
We have studied the properties of the EE in one-dimensional
disordered interacting systems. The EE saturates once the length of the segment
exceeds the localization length. We propose a heuristic expression which
describes the dependence of the EE on the segment length, system size
and localization length and compare it to numerical results. By fitting
the expression to the numerical data it is possible to extract the localization
length of an interacting system while calculating only its ground state 
many-body wavefunction.

\begin{acknowledgments}
Financial support from the Israel Science Foundation (Grant 686/10) is
gratefully acknowledged.
\end{acknowledgments}


\begin{thebibliography}{99}

\bibitem{anderson58} P.W. Anderson, Phys. Rev., {\bf 109}, 1492 (1958).

\bibitem{lee85}
For a review see:
P. A. Lee and T. V. Ramakrishnan, Rev. Mod. Phys. {\bf 57}, 287 (1985).

\bibitem{schmitt98} P. Schmitteckert , T. Schulze, C. Schuster, 
P. Schwab, and U. Eckern, Phys. Rev. Lett. 
{\bf 80}, 560 (1998); P. Schmitteckert, R. A. Jalabert, D. Weinmann, 
and J. L. Pichard,  Phys. Rev. Lett. {\bf 81}, 2308 (1998).

\bibitem{berkovits96} R. Berkovits and Y. Avishai, Phys. Rev. Lett. {\bf 76}, 
291 (1996).

\bibitem{carter05}
J. M. Carter and A.
MacKinnon, Phys. Rev. B {\bf 72}, 024208 (2005).

\bibitem{weiss07} Y. Weiss, M. Goldstein, and R. Berkovits
Phys. Rev. B {\bf 75}, 064209 (2007) .

\bibitem{amico08}
For recent reviews see:
L. Amico, R. Fazio, A. Osterloh, and V. Vedral, 
Rev. Mod. Phys. {\bf 80}, 517 (2008);
J. Eisert, M. Cramer, and M. B. Plenio,
Rev. Mod. Phys. {\bf 82}, 277 (2010); and references therein.

\bibitem{vojta06}
M. Vojta, Phil. Mag. \textbf{86}, 1807 (2006).

\bibitem{lehur08}
K. Le Hur, Ann. Phys. \textbf{323}, 2208 (2008).

\bibitem {goldstein11} M. Goldstein, Y. Gefen and  R. Berkovits, 
Phys. Rev. B {\bf 83}, 245112 (2011).

\bibitem{holzhey94}
C. Holzhey, F. Larsen, and F. Wilczek,
Nucl. Phys. B {\bf 424}, 443 (1994).

\bibitem{vidal03}
G. Vidal, J. I. Latorre, E. Rico, and A. Kitaev, 
Phys. Rev. Lett. {\bf 90}, 227902 (2003);
J. I. Latorre, E. Rico, and G. Vidal,
Quant. Inf. Comp. {\bf 4}, 048 (2004).

\bibitem{calabrese04}
P. Calabrese and J. Cardy,
J. Stat. Mech. P06002 (2004).

\bibitem{white92}
S. R. White \prl {\bf 69}, 2863 (1992);
\prb {\bf 48}, 10345 (1993).

\bibitem{dmrg}
U. Schollw\"{o}ck, Rev. Mod. Phys. \textbf{77}, 259 (2005);
K. A. Hallberg, Adv. Phys. \textbf{55}, 477 (2006).

\bibitem{korepin04} 
V. E. Korepin, Phys. Rev. Lett. {\bf 92} 096402 (2004).

\bibitem{affleck91} I. Affleck and A.W.W. Ludwig, Phys. Rev. Lett. 
{\bf 67}, 161 (1991).

\bibitem{igoli08}
F. Igloi and Y. C. Lin, J. Stat. Mech. P06004 (2008).

\bibitem{romer97}
R. A. R\"omer and M. Schreiber, Phys. Rev. Lett. {\bf 78}, 515 (1997).

\bibitem{apel_82} W. Apel, J. Phys. C {\bf 15}, 1973 (1982);
W. Apel and T. M. Rice, Phys. Rev. B 26, 7063 (1982);
T. Giamarchi and H. J. Schulz, Phys. Rev. B 37, 325
(1988).

\bibitem{g_formula} F. Woynarovich and H. P. Eckle, J. Phys. A {\bf 20}, L97 (1987); 
C. J. Hamer, G. R. W. Quispel, and M. T. Batchelor, ibid. {\bf 20}, 5677 (1987).

\bibitem{giamarchi}  T. Giamarchi, {\it Quantum Physics in One Dimension} (Oxford University Press, New York, 2003).

\bibitem{zhao06}
J. Zhao, I. Peschel, and X. Wang,
Phys. Rev. B {\bf 73}, 024417 (2006);

\bibitem{levine04}
G. C. Levine, Phys. Rev. Lett. {\bf 93} 266402 (2004);
G. C. Levine and D. J. Miller,
Phys. Rev. B {\bf 77}, 205119 (2008).

\bibitem{peschel05}
I. Peschel,
J. Phys. A: Math. Gen. {\bf 38}, 4327 (2005).

\bibitem{igloi09}
F. Igloi, Z. Szatmari, and Y.-C. Lin,
Phys. Rev. B {\bf 80}, 024405 (2009).

\bibitem{ren09}
J. Ren, S. Zhu, and X. Hao,
J. Phys. B {\bf 42} 015504 (2009).

\bibitem{affleck09}
I. Affleck, N. Laflorencie, and E. S. Sorensen,
J. Phys. A: Math. Theor. {\bf 42}, 504009 (2009). 

\bibitem{eisler10} V. Eisler and  I. Peschel,
Ann. Physik {\bf 522}, 679 (2010); V. Eisler and  S. S. Garmon,
Phys.Rev. B {\bf 82}, 174202 (2010).

\bibitem{kane92} C.L. Kane and M.P.A. Fisher, \prl \textbf{68}, 1220 (1992);
\prb \textbf{46}, R7268 (1992); \textbf{46}, 15233 (1992).

\bibitem{jia08}
X. Jia, A. R. Subramaniam, I. A. Gruzberg, and S. Chakravarty,
Phys. Rev. B {\bf 77}, 014208 (2008).

\bibitem{chakravarty10} S. Chakravarty, Int. J. of Mod. Phys. B
{\bf 24}, 1823 (2010)

\bibitem{anderson80}
P. W. Anderson, D. J. Thouless, E. Abrahams, and D. S. Fisher,
Phys. Rev. B {\bf 22}, 3519 (1980);
B. L. Altshuler V. E. Kravtsov, and I. V. Lerner, Zh.Eksp. Teor.
Fiz. {\bf 91}, 2276 (1986) [Sov. Phys. JETP {\bf 64}, 1352].


\end{thebibliography}
\end{document}